\documentstyle[amssymb,floats,tighten,prl,aps,epsf]{revtex}

\input epsf

\begin{document}


\def\Zsun{\thinspace\hbox{$\hbox{Z}_{\odot}$}}
\def\msun{\thinspace\hbox{$\hbox{M}_{\odot}$}}
\def\rsun{\thinspace\hbox{$\hbox{R}_{\odot}$}}
\def\lsun{\thinspace\hbox{$\hbox{L}_{\odot}$}} 
\def\.{\'{\i}}

\def\be{\begin{equation}} 
\def\ee{\end{equation}}

\title{ Strongest gravitational waves from neutrino oscillations 
at supernova core bounce}

\author{Herman J. Mosquera Cuesta$^{1,2,3}$ and Karen Fiuza$^{4}$  }

\address{{ $^1$Centro
Brasileiro de Pesquisas F\'{\i}sicas, Laborat\'orio de Cosmologia e
F\'{\i}sica Experimental de Altas Energias \\Rua Dr. Xavier Sigaud 150,
Cep 22290-180, Urca, Rio de Janeiro, RJ, Brazil --- e-mail: hermanjc@cbpf.br \\
$^2$Abdus Salam International Centre for Theoretical Physics,
Strada Costiera 11, Miramare 34014, Trieste, Italy \\  
$^3$Centro Latino-Americano de F\'{\i}sica, Avenida Wenceslau Braz 71, CEP 22290-140 Fundos, Botafogo, Rio de Janeiro, RJ, Brazil \\
$^4$Instituto de F\'{\i}sica - Universidade Federal do Rio 
Grande do Sul, Agronomia, Avenida Bento Gon\c calves 9500, Caixa Postal 
15051, Porto Alegre, RS, Brazil ::: kfiuza@if.ufrgs.br }}

\date{\today}

\maketitle

\begin{abstract}
Resonant  active-to-active  ($\nu_a \rightarrow  \nu_a$),  as well  as
active-to-sterile   ($\nu_a  \rightarrow   \nu_s$)   neutrino  ($\nu$)
oscillations  can take  place during  the core  bounce of  a supernova
collapse. Besides, over this phase, weak magnetism increases antineutrino
($\bar{\nu}$) mean  free paths, and thus its  luminosity.  Because the
oscillation feeds mass-energy into the target $\nu$ species, the large  mass-squared difference between species ($\nu_a  \rightarrow \nu_s$)  
implies a huge amount of energy to be given off as gravitational waves ($L_{\textrm{GWs}} \sim 10^{49}$~erg s$^{-1}$), due to anisotropic but 
coherent $\nu$ flow over the oscillation length. This asymmetric $\nu$-flux
is driven by both the spin-magnetic and the {\it universal spin-rotation}
coupling.  The novel contribution of  this paper stems from 1) the new
computation of the anisotropy parameter $\alpha \sim 0.1-0.01$, and 2)
the use of the tight constraints from neutrino experiments as  SNO and
KamLAND,   and    the   cosmic   probe   WMAP,    to   compute   the
gravitational-wave emission during neutrino oscillations in supernovae
core  collapse  and bounce.  We  show that  the  mass  of the  sterile
neutrino  $\nu_s$ that can  be resonantly  produced during  the flavor
conversions makes it a good  candidate for dark matter as suggested by
Fuller et {\it  al.} (2003).  The new spacetime  strain thus estimated
is  still several  orders of  magnitude larger  than those  from $\nu$
difussion (convection  and cooling)  or quadrupole moments  of neutron
star matter. This  new feature turns these bursts  the more promissing
supernova gravitational-wave signal that may be detected by observatories
as LIGO,  VIRGO, etc., for distances  far out to the  VIRGO cluster of
galaxies.

\vskip 0.8 truecm

\end{abstract}




\def\be{\begin{equation}}

\def\ee{\end{equation}}

\section{ Introduction} 

{\it  Supernovae  neutrinos  and  gravity waves.---}  That  outflowing
neutrinos ($\nu$s) from a  supernova (SN) generate gravitational waves
(GWs)  was firstly  pointed out  by  Epstein (1978). However,
over       the        first      $\sim       10$       milliseconds
(ms) (Mayle, Wilson \& Schramm 1987; Walker \& Schramm 1987) after  
the SN core  bounce the central
density gets  so high  that no radiation  nor even $\nu$s  can escape,
they are thus frozen-in and strongly coupled to the neutron matter 
($N^0$) as  described by the  Lagrangean (see Kusenko \& Postma 2002 
for this dynamics)

\be           
 L^{int}_{N^0\leftrightarrow            \nu}           =
\frac{G_F}{\sqrt{2}} \left[ \bar{N}^0    \gamma_\mu(1-\gamma_5)N^0 \right]
\left\{ \bar{\psi} \gamma^\mu(1-\gamma_5)\psi \right\}\;
,\label{interact} 
\ee

with  the $\nu$  field  $(\psi)$ satisfying  the time-dependent  Dirac
equation

\be  \left[i\gamma^0  \partial_0  + i\gamma^\alpha  \partial_\alpha  +
\rho(t)   v_\beta   \gamma^\beta   \left(\frac{1-\gamma_5}{2}\right)   
  - m_\nu\right] \psi = 0.
\label{dirac-mass}
\ee

At  this  phase  the  whole  proto-neutron  star  (PNS)  dynamics  is
dominated by gravity alone, and  can be appropriately described by the
general relativistic Oppenheimer-Volkoff equation for both the $N^0$ +
$\nu$ fluid (see Mosquera Cuesta 2002). As discussed by Mayle, Wilson
\& Schramm (1987); and Walker \& Schramm (1987),
it is over this early transient that most $\nu$ flavor conversions are
expected to  resonantly take place  and consequently the  super strong
GWs burst from the oscillation  process to be released.  GWs from this
decoupling  has  been suggested  to  likely  be  the ultimate  process
responsible for the neat kick given  to a nascent pulsar during the SN
collapse  (Mosquera Cuesta 2000; 2002).

The contention of this \textit{paper}  is a) to pave, in the framework
of  general   relativity  (GR),   the  pathway  to   this  fundamental
astrophysical process of generation  of GWs from $\nu$ oscillations in
a  PNS. b)  to demonstrate,  by taking  into account  experimental and
observational  constraints,  that $\nu$  oscillations  during SN  core
bounce do produce  GWs of the sort predicted  by Einstein's GR theory,
and more  crucial yet,  c) to  stress that these  bursts are  the more
likely SN GWs-signals to  be detected by interferometric observatories
as LIGO, VIRGO, GEO-600, etc.  We speculate that such a signal perhaps
might have  been detected  during the SN1987a  event, despite  the low
sensitivity of the detectors at the time. Some claims in this direction 
were presented by Aglietta, Amaldi, Pizzella, et al.\cite{amaldi89}, and
related papers.

\section{The mechanism for generating GWs during neutrino oscillations}

To start  with, let us recall  how the production of  GWs during $\nu$
oscillations proceeds by considering  the case of oscillations between
active  and sterile  neutrinos in  the supernova  core.  The essential
point  here  is  that   oscillations  into  sterile  neutrinos  change
dramatically   the   energy   and   momentum  (linear   and   angular)
configuration of the system:  neutrinos plus neutron matter inside the
PNS (check  Eq.(\ref{dirac-mass})). In particular,  flavor conversions
into sterile neutrinos drive a large mass and energy loss from the PNS
because once they  are produced they freely escape  from the star. The
reason: they  do not interact  with any ordinary matter  around, i.e.,
they do  couple to active  $\nu$ species but  neither to $Z^0$  nor to
$W^\pm$ vector bosons.  This means that oscillations into steriles, in
dense  matter, take  place  over longer  oscillation lengths, compared 
to $\nu_a \rightarrow  \nu_a$, and  the
steriles  encounter infinite mean  free paths  thereafter. Physically,
the  potential, $V_s(x)$,  for sterile  neutrinos in  dense  matter is
zero. In addition, their probability of reconversion, still inside the
star, into active species is  quite small (see discussion below). This
outflow translates into a noticeable  modification of the PNS mass and
energy quadrupole distribution, which  as discussed below is dominated
from the very beginning by rotational and magnetic field effects.

Since most  steriles neutrinos escape along the  directions defined by
the    dipole    field    and    angular   momentum    vectors    (see
Fig.\ref{4-dupole}), then the $\nu$ outflow is at least quadrupolar in
nature. This produces a super strong gravitational-wave burst once the
flavor conversions  take place,  the energy of  which stems  from the
energy and momentum of the  total number of neutrinos participating in
the oscillation process\footnote{The attentive reader must regard that
neutrinos carry away almost all of the binding energy of the just-born
neutron  star,  i.e.,   $\Delta  E_\nu  \sim  3\times  10^{53}$~erg.}.
Further, the  gravitational-wave signal  generated this way  must 
exhibit  a  waveform with a  {\it Christodolou's memory}  
(Mosquera Cuesta 2002).

The remaining configuration  of the star must also  reflect this loss.
Hence, its own matter and energy distribution becomes also quadrupolar.
Because  this quadrupole  configuration (the  matter and  energy still
trapped  inside the just-born  neutron star)  keeps changing  over the
time scale for which most of the oscillations take place, then GWs must
be  emitted  from the  star  over that  transient.   At  the end,  the
probability  of conversion  and  the $\nu$  flux anisotropy  parameter
($\alpha$, see below)  determine both how much energy  partakes in the
process  and  the  degree  of  asymmetry  during  the  emission.  Both
characteristics are determine next.


\vskip 0.5 truecm

EDITOR PLACE  FIGURE 1 AROUND HERE  !!!!!

\vskip 0.5 truecm

The case for oscillations among active species is a bit different, the
key feature being that mass and energy is relocated from one region to
another inside the PNS, especially because of the {\it weak magnetism}
of antineutrinos that allows them  to have larger mean free paths (and
thus oscillation lengths)  (Horowitz 2002).  In addition, oscillations
of electron  neutrinos into muon  or tauon neutrinos leave  these last
species  outside their  own  neutrinospheres, and  hence  they are  in
principle free to stream away. These neutral-current interacting $\nu$
species must  be the very first  constituents of the  $\nu$ burst from
any supernova  since most $\nu_e$s are essentially  trapped. This must
also  generate GWs during  that sort  of flavor  conversions, although
their  specific strength  (strain) must  be  a bit  lower compared  to
conversions into sterile neutrinos  where almost all the $\nu$ species
may participate, and the large $\Delta m^2$ in the process.

Since the sterile neutrinos escape the  core over a time scale of a few
ms, the number  of neutrinos escaping and their angular
distribution  is  sensitive   to  the  instantaneous  distribution  of
neutrino  production sites.   Since thermalization  cannot  occur over
such short times ($\Delta T^{\rm  thermal} \sim 0.5$~s), and since the
neutrino production rate is sensitive  to the local temperature at the
production  site, the  inhomogeneities during  the collapse  phase get
reflected in  the inhomogeneities in the escaping  neutrino fluxes and
their  distributions. Because  of  both the  $\nu$ spin-magnetic field
($\vec{B}$)  and $\nu$ spin-angular momentum ($\vec{J}$) coupling 
the asymmetries  in these distributions
can give rise to quadrupole moments, which must generate gravitational
waves as suggested by Mosquera Cuesta (2000; 2002), and dipole moments
which can explain the origin of pulsar kicks (Kusenko \& Segr\`e 1997; 
Fuller et {\it al.} 2003).

Fixed by  the {\it probability of  oscillation}, $P_{\nu_a \rightarrow
\nu_a}^{\nu_a \rightarrow
\nu_s}$, the {\it fraction} of  neutrinos that can escape in the first
few milliseconds is, however, {\it small}. Firstly, the neutrinos have
to be produced roughly within  one mean free path from their resonance
surface.  Secondly,  since in the case of  ${\nu_a \rightarrow \nu_s}$
oscillations $m_s$ is  the heaviest neutrino species, the  sign of the
effective potential  $A(x)$ (see  discussion below) and  the resonance
condition   indicates  that   only  ${\nu}_e$s and  the  antineutrinos 
$\bar{\nu}_\mu$ and $\bar{\nu}_\tau$  can undergo resonant conversions.
In  Section  IV  we  address  all  these  issues  and  determine  this
fundamental  property of the  mechanism for  producing GWs  from $\nu$
flavor conversions.

\section{Anisotropic neutrino outflow: Origin and computation}

To provide a physical foundation  for the procedure introduced here to
determine  the  neutrino {\it  asymmetry  parameter}, $\alpha$,  which
measures how large is the deviation of the $\nu$ flux from a spherical
one, we recall next two  fundamental effects that run into action once
a PNS  is forming after the  supernova collapse. We  stress that other
physical  process such  as convection,  thermalization, etc.,  are not
relevant over  the time scale under consideration:  $\Delta T_{\rm osc}
\sim  3-10$~ms after the  SN core  bounce. Those  effects take  a more
longer time ($\sim 30-100$~ms) to start to dominate the physics of the
PNS, and  therefore do  not modify in  a sensitive manner  the picture
described  below\footnote{ At this point, a note of warning regarding 
the time scale we are using for the present calculations is of worth. 
This is specially so in the light of the very recent paper by Loveridge
\cite{loveridge}, where a very important extension of our original idea,
introduced in Refs.\cite{herman2001}, is detailedly provided. In his 
computation of the gravitational radiation emission from an off-centered 
flavor-changing $\nu$ beam Loveridge used a time scale of $\Delta 
T^{\nu-\rm burst} \sim 10$~s, apparently based on the duration of the 
$\nu$-burst from SN1987A. 

The great novelty in Loveridge's paper is the prediction of a periodic
GWs signal from the flavor-changing $\nu$ beam eccentrically outflowing
from the just-born pulsar. This GWs signal would have characteristics
that make it observable by both LIGO and LISA GWs interferometers.
Interestingly in itself, various arguments favouring evidence for
regularly pulsed neutrino emission from SN1987a in the period range
between (8.9-11.2)~s were given by Harwit, et al.\cite{harwit87} and 
by Saha and Chattopadhyay \cite{saha91}. However, Fischer checked
systematically for periodicities between (5-15)~ms in the $\nu$ burst
from  SN1987A \cite{fischer87}.  He disclaimed all those hypotesis by
showing that the multitude of medriocre period fits seemed to be rather
typical for events distributed randomly instead of periodically
\cite{fischer87}.  Thus no significant periodicity exists in the
arrival time of the neutrinos from SN1987A as detected by Kamiokande 
II and IMB detectors.  Nonetheless, any evidence for such a regular 
$\nu$ signal in a forthcoming (future) supernova might decidedly favor
Loveridge's GWs mechanism from $\nu$ oscillations in nascent pulsars.

As it stands, however, Loveridge's mechanism is decidedly different
from ours in several respects. Firstly, we do not invoke an
off-centered rotating $\nu$ beam for producing the GWs emission from
the $\nu$ oscillations.  In our mechanism the neutrino outflow is
simultaneously acted upon by both the pulsar centered magnetic field
and angular momentum vectors, as we describe below. Thus, the
$\nu$-spin coupling to both vectors turns out to be the source of the,
at least, quadrupolar $\nu$ outflow and GWs emission during the flavor
conversions. Secondly, as a consequence of this $\nu$ escape from the
star the resulting GWs signal, in principle, is not periodic, as
opposed to Loveridge's. Indeed, the GWs signal will look much like the
one computed by Burrows and Hayes \cite{burrows96}, including the
appearance of a Christodoulou's memory in the waveform.  Thirdly, as
discussed next, the overall time scale for the process to take place in
our mechanism is about three orders of magnitude shorter compared to
one assumed by Loveridge.

The $\nu$ oscillation time scale and the duration of the GWs emission
are both crucial features of both the mechanisms above discussed. In
this regards, we advise that a very extensive set of references (here
we just quote a few of them) showed that the $\nu$ signal from SN1987A
exhibits a peculiar time profile \cite{sato87,arnett87,cowsik88}.
According to Refs.  \cite{sato87,arnett87,cowsik88}, the $\nu$ burst
observed from SN1987A is bunched into three clusters around (0.0-0.107)
~s, (1.541-1.728)~s, (9.219-12.349)~s. In particular, Cowsik
\cite{cowsik88} claimed that during the very early phase the lighter of
the neutrinos arrived and by $\Delta T \sim 0.1$~s all of the neutrinos
above the IMB threshold of 20~MeV had already gone past and thus were
not seen by this $\nu$ detector.  Therefore, if one stands on these
pieces of evidence it is clear that the largest portion of the total
$\nu$s from SN1987A were emitted over a time scale smaller than 100~ms.
This last time scale is an order of magnitude larger than the one we
are favouring here, one which is taken from most of the theoretical
analysis and numerical simulations of supernova core collapse and 
$\nu$ emission (see for instance Refs.  10,11,12 in Sato and Suzuki
\cite{sato87}; and also Burrows and Lattimer \cite{burrows86}; and
Refs. \cite{schramm87A,schramm87}). A possible explanation of such a
difference may be that a large portion of the released electron $\nu$s
undergone flavor conversions into sterile $\nu$s over a time scale much
shorter than 100~ms, a reason of why they went undetected.

In brief, although from the observational point of view, we may agree
that the ($\sim 10$~s) $\nu$ emission time scale assumed by Loveridge
\cite{loveridge} appears a reasonable one, we think of it is not a very
realistic one to picture out the $\nu$ flavor conversion mechanism
inside the nascent neutron star, specially if one takes into account
that the $\nu$ oscillation process implies a very short time scale: 
the resonance or {\it coherence length} time scale. As claimed above, 
this time scale must be related to the {\it oscillation length} (not 
the system's response time scale) over which most of the conversions 
must take place. Moreover, oscillations of massive neutrinos are 
damped when the propagation distance is greater than the coherence 
length

\be
L_{\rm coh} = \lambda_{\rm osc} \left(\frac{E}{\Delta E}\right) \; .
\ee

Here $E$ and $\Delta E$ are, respectively, the $\nu$ energy and energy
spread determined by the production and detection conditions.  In a
supernova, the neutrinos nonforward scatter in continuous energy
distributions so that ${\Delta E} \simeq E$, and hence the coherence
length is nearly the oscillation length \cite{pantaleone92}, which
fixes the time scale $\Delta T_{\rm osc}$ we call for henceforth. Yet,
in the early phases of a SN the neutrino flux is so large that the
weak-interaction potential created by the neutrinos is compatible to
that of the baryon matter around. Thus, neutrinos can be thought of as
a dominant background medium that acts as a coherent superposition of
flavor states that drives the conversions in a nonlinear way. In other
words, the oscillations become ``synchronized", which means that all
modes oscillate {\it collectively} with the {\it same frequency}
\cite{samuel93,raffelt02}. Such a frequency should be related to the
oscillation or coherence length, and through it to the oscillation time
scale. Thus, this last behaviour adds to our argument in favour of a
shorter time scale for the overall oscillations to take place. So, the
$\Delta T_{\rm osc} \lesssim 10$~ms is well-fundamented.  Besides, a
time scale that long as Loveridge's $\sim 10$~s strongly disagree with
Spruit and Phinney constraint on the overall time scale:  $\sim
0.32$~s, for the kick driving mechanism \cite{phinney98}.}.  Indeed, if
$\nu$ thermalization, for instance, already took place, then the
oscillations are severely precluded since oscillations benefit of the
existence of energy, matter density and entropy gradients inside the
PNS (Bilenky, Giunti and Grimus 1999; Akhmedov 1999), which are
``washed out'' once thermalization onsets.


\vskip 0.5 truecm

  EDITOR PLEASE PLACE TABLE 1 AROUND HERE !!!  

\vskip 0.5 truecm

\subsection{Why no room for $\nu$-driven convection over 
$\Delta T_{\rm osc}$ }

In  order  to  back  the  dismissal  in  our  discussion  on  neutrino
oscillations  of the  effects of  convection inside  the proto-neutron
star, we would  like to take advantage of  some arguments presented in
the  {\it state-of-the-art}  of the  subject by  Janka,  Kifonidis  \&
Rampp  (JKR, 2001),  who provided  detailed analysis  about convection
inside the  nascent neutron star. In particular,  these authors showed
that  the  growth time scale  of  convective instabilities  ($\tau_{\rm
cv}$)  in  the neutrino-heated  region  (adjacent  layers outside  the
just-born  neutron   star,  of  relevance   for  successful  supernova
explosions)  depends on  the gradients  of entropy  and  lepton number
through the growth rate of Ledoux convection, $\sigma_L$, as

\be \tau_{\rm  cv} \simeq \frac{\ln(100)}{\sigma_L}  \simeq 4.6 \left[
\frac{g} {\rho}  \left(\frac{\partial \rho}{\partial s}\right)_{Y_e,P}
\frac{ds}{dr} +  \left(\frac{\partial \rho}{\partial Y_e}\right)_{s,P}
\frac{dY_e}{dr} \right]^{-1/2} \; , \ee

or equivalently

\be  \tau_{\rm  cv}  \simeq  20~{\rm  ms}  \left(\frac{R_s}{R_g}  -  1
\right)^{1/2}        \frac{R_{g,7}^{3/2}}{\sqrt{M_1}}       \;       ;
\label{convec-time} \ee

where $M_1 = 1 M_\odot$, and $R_s$ and $R_g$ define the shock and gain
radius, respectively, and $R_{g,7}$ is  a function of the gain radius,
the  temperature inside  the  star, and  the  neutrino luminosity  and
energy.  Here the estimates were obtained for  $g = GM/R_g^2$,
$(\partial  \rho/\partial  s)_P  \sim   -\rho/s$,  and  $  ds/dr  \sim
-\frac{1}{2}  \frac{s}{(R_s  -  R_g)}$  (see JKR  2001,  for  further
details).

Numerical   simulations  demonstrate   that   convection  inside   the
proto-neutron star does  start as early as a  few tens of milliseconds
after  core bounce. It  develops in  both ($i$)  unstable surface-near
regions, i.e.,  in layers around  the neutrinosphere where  density is
$\rho  \leq  10^{12}$~g~cm$^{-3}$,  and  ($ii$) deeper  layers  of
density $\rho  \geq 10^{12}$~g~cm$^{-3}$), where  a negative lepton
number  gradient appears.   Despite  of this  piece  of evidence,  the
time scale  defined  by  Eq.(\ref{convec-time})  is  relatively  long
compared to  both the estimated time interval  for the deleptonization
process to take place: $\Delta  T_{\rm delept} \sim 10$~ms, i.e., time
over  which most  electron neutrinos  are produced  (Mayle,  Wilson \&
Schramm 1987; Walker \& Schramm  1987), and the core bounce time scale:
$\Delta  T_{\rm bounce}  \sim  20$~ms;  where the  large  part of  the
neutrino luminosity associated with  other flavors is produced through
processes like  bremsstrahlung, neutrino-neutrino and neutrino-nucleon
scattering within less than 5~ms (Mayle, Wilson \& Schramm 1987; Walker 
\& Schramm 1987; JKR
2001,  and references  therein). Indeed,  from the  convective regions
below the neutrinosphere  neutron fingers dig into the  star and reach
its  center in  about  one  second. Then  they  propagate outwards  to
englobe almost  all the exploding star. Under  the physical conditions
dominant over  the first  10-20 ms after  core bounce one  expect most
neutrino  oscillations   of  all  flavors  to  take   place  at  that
time. Thence,  convective effects are not relevant  during a time scale
that short.   As such,  it cannot modify  in a significant  fashion our
analysis  regarding   the   mechanism   for  the   generation   of
gravitational waves from neutrino oscillations, here highlighted.

Moreover, JKR (2001) stressed the ``desastrous'' r\^ole of rotation for
convection.  A  high rotation velocity  of the just-born  neutron star
reduces  dramatically  the  effects   of  convection  because  of  the
suppression  of  the   neutrino-nucleon  interaction  due  to  nucleon
correlations in  the nuclear medium composing  the proto-neutron star.
Physically, rotation leads to a suppression of convective motions near
the rotation axis because of  a stabilizing stratification of the star
matter specific angular  momentum. In passing, we stress that a similar 
effect is
also expected  from the action  of a background magnetic  field.  Both
effects, rotation and  magnetic field, then appear to  be more crucial
for the physics of neutrino interactions inside the newly-born neutron
star, and for the production  of GWs during the oscillation transient.
Thus we address both of them next.

\subsection{$\nu$-rotation interaction}
 
That gravity couples to neutrinos is well-known since Dirac.  The very
first work, as far as we are  aware of, to show that the particle spin
and PNS rotation   couples    gravitationally, a    {\it   universal}
feature (Soares \& Tiomno 1996), was that  of Unruh (1973). It was
shown that  a consistent  minimal-coupling generalization to  a curved
background (a Kerr  spacetime in that case) of  the $\nu$ equations is
possible,  and that  it leads  to  equations {\it  separable} for  the
radial and  angular components, though coupled. For  a massless $\nu$
field (compare to  Eq.(\ref{dirac-mass})), and standard Minkowski space
Dirac matrices $\gamma^A$, the Dirac equation derived by Unruh reads


\be \gamma^A \left(\frac{\partial}{\partial x^A} - \Gamma_A \right)\psi 
= 0\; , \; \; \; \label{dirac-kerr}
\ee

where the Dirac $\gamma^A$ matrices relate to the Kerr spacetime metric
through

\be \gamma^A \gamma^B + \gamma^B \gamma^A = 2 g_{\rm Kerr}^{A B}\; .
\label{gammas-clifford-algebra}
\ee

Eq.(\ref{gammas-clifford-algebra}) shows  that the $\gamma^A$ matrices
satisfy  the Clifford's  algebra. Further,  the  spin-affine connections
$\Gamma_A$  in Eq.(\ref{dirac-kerr})  are uniquely  determined  by the
relations

\be   \Gamma_A   \gamma^B  -   \gamma^B   \Gamma_A  =   \frac{\partial
\gamma^B}{\partial x^A} + \Gamma_{\alpha A}{^B\gamma^\alpha} \;, \; \;
{\rm and} \; \; \; \; {\rm tr} (\Gamma_A) = 0\; .  \ee

From  the  $\nu$-number  current,  $J^A(x)  =  \bar{\psi}(x)  \gamma^A
\psi(x)$,  it  was  shown  that  the $\nu$-number  density  is  always
positive and  given as $ (-g)^{1/2} J^t(x)  = (-g)^{1/2} \bar{\psi}(x)
\gamma^t \psi(x)$  (see Unruh 1973 for  details). The complete
analysis of the coupling shows that the $\nu$ field in this background
is not  {\it superradiant},  as opposed to  the case of  the classical
fields studied previously.

Vilenkin (1978) extended the above analysis and showed that,
upon admitting {\it helicity} ($L$) to be a {\it good} quantum number,
the angular distribution $F_{j m} (\theta)$ of the thermal fermion gas
of $\nu$s  ($L=+1$) and  $\bar{\nu}$s ($L=-1$) in  the mode  $(j, m)$,
with  the function  $F_{j  m} (\theta)$  satisfying the  normalization
condition $2\pi  \int_0^\pi F_{j  m}(\theta) \sin\theta d\theta  = 1$,
leads  to an  {\it  asymmetric} $\nu$  emission  (see illustration  in
Fig.\ref{4-dupole}) from  a Kerr black hole (BH),  of specific angular
momentum $a \equiv J/M$, described by

\be  \frac{dN}{dt  d\omega d\theta}  =  \frac{1}{8\pi^2} M^2  \omega^2
\frac{\sum_\pm (1 \pm L a \cos\theta)}{ \left\{e^{\left[\frac{2\pi}{k}
(\omega \pm \frac{1}{2}\Omega_{BH}) \right]} + 1\right\} } \;.
\label{distrib}
\ee

Eq.(\ref{distrib})  shows that  more $\bar{\nu}$s  are emitted  in the
direction parallel to the BH's  spin, whilst more $\nu$s escape in the
antiparallel direction.{\footnote{Notice  that this behaviour  is also
manifest  in  the  case  of  the magnetic  field  $\nu$-spin  coupling
discussed  below. Therefore, these effects affect the oscillation 
probability, as we discuss later on.}  Further,   for  other  
weak-interacting  particles emitted  from the  BH {\it  parity} is  
{\it not  conserved} (Vilenkin 1978).  More fundamental yet, Vilenkin 
(1978) work demonstrates that the same
physics must be valid for any other rotating star. In other words, the
$\nu$ spin coupling  to rotation, in a gravitational  background, is a
{\it  universal} feature,  regardless which  the spacetime  source can
be. Here onwards we shall take  advantage of this feature for the case
of neutrino  emission from a rotating  NS, and suggest  that the basic
quadrupole nature  of the  $\nu$ emission from  the PNS stems  in part
from this  spacetime effect. The  other fundamental effect  we address
right next.


\vskip 0.5 truecm

  EDITOR PLEASE PLACE FIGURE 2 AROUND HERE !!!  

\vskip 0.5 truecm



\subsection{$\nu$-$B$ field interaction}
 
That the electromagnetic properties of  $\nu$s are modified due to its
interaction  with a  background  matter distribution  is a  well-known
fact.   This  is  reflected  in  the additional  contribution  to  its
self-energy  stemming from the  $\nu$ spin-to-magnetic  field coupling
$(\vec{B} \cdot \vec{k}_\nu)$.  In  the case of $\nu_{\tau} \to \nu_e$
oscillations,  for  instance,  the  geometry of  the  $\nu$-sphere  is
dramatically deformed by $\vec{B}$ (see Fig.\ref{4-dupole}), an effect
that strongly  depends upon the  relative pointing directions  of both
$\vec{B}$  and $\nu$  momentum $\vec{k_\nu}$.   The  magnetic coupling
distorts the $\nu_\tau$-surface (``sphere'') in  such a way that it is
no more concentric with the $\nu_e$-sphere (see Fig.1 and 2 in Kusenko
1999).  Therefore,  $\nu_\tau$s  escaping  paralelly-pointing  to  the
$\vec{B}$ field have a {\it lower} temperature than those flowing away
in the opposite (anti-parallel) direction.


At   the   same  time,   the   $\nu$   spin   coupling  to   rotation,
Eq.(\ref{distrib}) (Vilenkin 1978),  drives also an effective momentum
(and thus energy flux)  asymmetry along the angular momentum direction
$\vec{J}$, as shown in Fig.\ref{4-dupole}.  For a relative orientation
$\theta(\vec{B} \leftrightarrow \vec{J}) \neq 0$ between $\vec{B}$ and
$\vec{J}$,  i. e.,  a canonical  pulsar, this  combined action  on the
escaping $\nu$s of a rotating background spacetime plus magnetic field
makes   their  $\nu$-sphere  a   decidedly  distorted   surface.  More
precisely,  the  volumetric region  obtained  by  rotating around  the
hatched regions in Fig.\ref{4-dupole},  becomes at least a quadrupolar
outflowing energy distribution.  This is the source of  the strong GWs
bursts  in  this  mechanism  when  the  oscillations  ensue.  It  also
justifies the  {\it large value} of the  anisotropy parameter $\alpha$
used in Eq.(14) below (see also Burrows \& Hayes 1996).


\vskip 0.5 truecm

  EDITOR PLEASE PLACE TABLE 2 AROUND HERE !!!  

\vskip 0.5 truecm

\subsection{The anisotropy parameter}

Based  on the  concommitant  action of  both  effects: the  $\nu$-spin
coupling to both the magnetic field and rotation described previously,
one   can  determine   the   $\nu$  flow   anisotropy   in  a   novel,
self-consistent fashion by defining $\alpha$ as the ratio of the total
volume  filled   by  the  distorted  $\nu$-spheres  to   that  of  the
proto-neutron   star   ({\rm   PNS}),    as   one   can   infer   from
Fig.\ref{4-dupole}.   The  $\nu_e$-sphere  radius  of  a  non-magnetic
non-rotating star is obtained from the condition:

\be       
\tau_{\nu_e}(R_{\nu_e})       =      \int_{R_{\nu_e}}^\infty
{\mathcal{K}}_{\nu_e} ~ \rho(r) ~ dr = \frac{2}{3} \;, 
\ee

where $\tau_{\nu_e}$ is the optical depth, and ${\mathcal{K}}_{\nu_e}$ the
scattering  opacity  for electron  neutrinos,  and  $\rho$ the  matter
density.  Following  Burrows, Hayes \& Fryxell (1995)  one  can  take  
hereafter
$R_{\nu_e}  \equiv  R^{\nu_e}_{\rm PNS}  \equiv  R_{\nu} \sim  35$~km,
which  is  of  the  order  of  magnitude  of  the  oscillation  length
$\lambda_{\nu}$ of  a typical supernova $\nu$, constituent  as well of
the  atmospheric $\nu$s  for  which ${\Delta  m^2}  \sim 10^{-3}  {\rm
eV}^2$ has been estimated by Superkamiokande $\nu$ detector (Fukuda et
{\it al.} 1998)

\be 
\lambda_{\nu} \sim 31~{\rm  km}~ \left[\frac{ E_{\nu_e}}{10 ~{\rm MeV}
}\right] \left(\frac{ 10^{-3} ~{\rm eV}^2 }{\Delta m^2}\right)\; .  
\ee

Therefore, resonant conversions between  active species may take place
at the position $r$ from the center defined by 

\be r  = R_{\nu_e} + \delta_0 \cos  \phi\; , \label{nu-sphere-general}
\ee

with  $\phi$ the  angle between  the $\nu$-spin  and  $\vec{B}$, i.e.,
$\cos   \phi   =   \frac{(\vec{k}  \cdot   \vec{B})}{\vec{k}}$, and 

\be \delta_0 = \frac{e \mu_e  B}{2 \pi^2 (dN_e/dr)} \sim 1-10~\rm km\;
, \ee

for $B \sim 10^{14-15}$~G, respectively. Here $e$, $N_e = Y_e N_n$ and
$\mu_e$ represent the electron charge, density and chemical  potential,
respectively.   This  defines  in  Fig.\ref{4-dupole}  an  ellipsoidal
figure of equilibrium with semi axes

\be a  = R_\nu  + \delta_0; \hskip  0.5 truecm  {\rm and }  \hskip 0.5
truecm b = R_\nu, \ee

and volume (after rotating around $\vec{B}$):

\be V_{ellips.} = \frac{4}{3}\pi R^2_\nu (R_\nu + \delta_0) \; .
\label{ellipsoidal-volume}
\ee

Meanwhile,   the  $\nu$-spin   coupling  to   rotation   described  by
Eq.(\ref{distrib}) generates an asymmetric lemniscate-like plane curve
(see Fig.\ref{4-dupole})

\be  r = R_\nu~(1  \pm L~a~\cos~\theta)\;  , \label{lemniscate-sphere}
\ee

which upon  a $2 \pi$ rotation  around the star  angular momentum axis
generates a volume:

\be  V_{lemnisc.} =  \frac{1}{4}(2  R_\nu)^2 \times  2  \pi \bar{y}  +
\frac{1}{4}(R_\nu)^2 \times 2 \pi \bar{y}\; , \ee

where  the  quantity  $\bar{y}$  (and  $\bar{x}$) is  defined  as  the
location of the centroid of one of the lobes of that plane figure with
respect  to its  coordenate center  (x,y),  and is  obtained from  the
standard  definition $\bar{y} =  \frac{\int y  dA}{\int dA}$.  After a
long, but straightforward, calculation one obtains

\be \bar{y}  = \frac{\sqrt{\pi}}{\Gamma^2(1/4) }~R_\nu\;  , \hskip 0.5
truecm \bar{x} = \frac{4 \sqrt{\pi}}{\Gamma^2(1/4) }~R_\nu \; .  \ee

Thence, for  a PNS as  the one modeled  by Burrows, Hayes  \& Frixell
(1995), with parameters as given in Table \ref{parameters-PNS}
one obtains


\be  \alpha_{\rm min} =  \frac{ V_{ellips.}  + V_{lemnisc.}  }{ V_{\rm
PNS} } \sim 0.11-0.01\; , \label{param-alpha} \ee

a figure clearly compatible  with that one in Burrows \& Hayes (1996). As
such, this is essentially  a new result of this paper. The attentive
reader   must    notice   in   passing   that    the   definition   in
Eq.(\ref{param-alpha}) does  take into account  all of the  physics of
the neutrino  oscillations: luminosity, density  gradients and angular
propagation,                                                      since
Eqs.(\ref{nu-sphere-general},\ref{lemniscate-sphere})  do  gather  the
relevant information regarding the  spatial configuration of the $\nu$
luminosity  in  as much  as  is done  in  the  standard definition  of
$\alpha$ (Burrows \& Hayes 1996;  M\"uller \& Janka 1997)

\be \alpha(t)  \equiv \frac{1}{L_\nu (t)} \; \int_{4  \pi} d\Omega' \;
\Psi(\theta',\phi') \; \frac{L_\nu (\Omega',t)}{d\Omega'} \; .
\label{alpha-standard}
\ee

Indeed, one can get the ``flavour'' of the relationship between these two
definitions by  noticing that the quantity $L_\nu  (\Omega',t)$ in the
integrand  of  Eq.(\ref{alpha-standard}) can  be  expressed as  $L_\nu
(\Omega',t)   \equiv  L_\nu(t)  F(\Omega')   $,  where   the  function
$F(\Omega')$ contains  now all  the information regarding  the angular
distribution  of  the neutrino  emission.  Hence  $L_\nu(t)  $ can  be
factorized    out     of    the    integral     and    dropped    from
Eq.(\ref{alpha-standard}). This  converts Eq.(\ref{alpha-standard}) in
a relationship  among ({\it solid}) angular  quantities, which clearly
can be reduced  to a volumetric one, similar to  the one introduced in
Eq.(\ref{param-alpha}), upon a  transformation using the definition of
{\it  solid angle}  in the  form of  {\it Lambert's  law}:  $d\Omega =
\frac{dA \cos\theta}{R^2}$, and applied to the sphere representing the
PNS. Here $\theta$, measured from  a coordinate system centered at the
PNS (source  coordinate system in M\"uller \&  Janka 1997),  plays the
role of the  angle between the direction towards  the observer and the
direction  $\Omega'$  of  the   radiation  emission  in  Eq.(24),  and
subsequents, in M\"uller \&  Janka (1997). Therefore, the novel result
here presented is physically consistent  with the standard one for the
asymmetry parameter $\alpha$.



\section{Enlarged $\nu$ and GWs luminosity from oscillations in dense 
matter}

$\nu$ outflow  from a  SN core  bounce is a  well-known source  of GWs
( Epstein 1978; Burrows \& Hayes 1996; M\"{u}ller \& Janka 1997;
Mosquera Cuesta 2000; 2002).  Numerical  simulations (M\"{u}ller \& 
Janka 1997) showed that, in  general, the  fraction of  the total
binding energy emitted as GWs by pure $\nu$ convection is: $E^\nu_{\rm
GW}  \sim$ [10$^{-10}$-10$^{-13}$]~M$_\odot$c$^2$,  for a  $\nu$ total
luminosity: $L_{\nu} \sim 10^{53}$ erg~s$^{-1}$.

Unlike  GWs  produced  by  $\nu$ convection  (M\"{u}ller \& Janka 1997), 
in  the production  of  GWs  via  $\nu$  oscillations (Mosquera Cuesta 
2000; 2002)  ($\nu_a
\leftrightarrow \nu_a$ or  $\nu_a \leftrightarrow \nu_s$) there exists
two  main  reasons  for  expecting  a major  enhancement  in  the  GWs
luminosity  during the  transition:  a) the  conversion itself,  which
makes the overall luminosity  ($L_{\nu_x}$) of a given $\nu_x$ species
to  grow   by  a  large   factor:  $\Delta  L_{\nu_x}   \leq  10\%
L^{\textrm{total}}_\nu $,  see below.  The enhancement stems  from the
mass-energy being  given to,  or drained from,  the new  $\nu$ species
into which oscillations take place. This augment gets reflected in the
species  mass-squared  difference, $\Delta  m^2$, and  their  relative
abundances: one  species is number-depleted  while the other  gets its
number  enhanced. But, even  if the  energy increase,  or give  up, is
small, b) the abrupt resonant conversion over the transition time 
(see Table \ref{time-scales})

\be \Delta  T_{\textrm{osc}} \equiv \frac{  \lambda_{\textrm{osc}} } {
\bar{V}_{\nu-Diff.} } \sim  [10^{-4}-10^{-3}]~\rm s , 
\label{oscillation-time}
\ee

also magnifies  transiently $L_{\nu_x}$. Here $\lambda_{\textrm{osc}}$
defines the oscillation length (computed below), and $\bar{V}_{\nu-Diff.} 
\sim
10^9$~cms$^{-1}$    the    convective    $\nu$   diffusion    velocity
(M\"{u}ller \& Janka 1997). In Section IV we estimate the transition 
probability:
$P_{\nu_a   \rightarrow   \nu_a}   (|\vec{x}-\vec{x}_0|)$,   $P_{\nu_a
\rightarrow  \nu_s}(|\vec{x}-\vec{x}_0|)$, the quantity  that measures
how many $\nu$s can indeed  oscillate. This probability also fixes the
total  amount of  energy participating  in the  generation of  the GWs
through this mechanism, as shown in Section V.

If flavor  transitions can indeed  take place during  supernovae (SNe)
core collapse  and bounce, then they  must leave some  imprints in the
SNe  neutrino spectrum.  Main observational  consequences  of neutrino
conversions inside SNe include a) the partial or total disappearance of
the  neutronization  peak; the moment at which  most  $\nu_e$s  are 
 produced, b)  the
interchange  of the  original spectrum  and the  appearance of  a hard
$\nu_e$  spectrum, together  with c) distorsions of  the $\nu_e$  energy
spectrum  and d) alterations of  the  $\bar{\nu}_e$  spectrum (Dighe  \&
Smirnov 2000). As discussed below  in Section 4.4, observations of the
neutrino burst  from SN1987a have allowed  to put some  bounds on both
$\nu_a  \rightarrow  \nu_a$ and  $\nu_a  \rightarrow \nu_s$ classes of
flavor conversions.

\subsection{Resonance, adiabaticity and the role of weak magnetism}

As argued by Mosquera  Cuesta (2000; 2002) $\nu$ oscillations in vacuum
produce  no GWs. In  the case  of active-to-active  $\nu$ oscillations
(essentially   the    same   physical   argument    holds   also   for
active-to-sterile  $\nu$  oscillations),  the  main  reason  for  this
negative result is that this class of conversions do not increase in a
significant  figure the total  number of  particles escaping  from the
proto-neutron star. In the case of active-to-active $\nu$ oscillations
in dense matter,  the process generates no GWs  since the oscillations
develop with the neutrinos having  very short mean free paths, so that
they motion outwards can be envisioned as a standard difussion process.

However, if one takes into account the novel result by Horowitz (2002)
the  situation may  change  dramatically.  According  to this  author,
because of the active  {\it antineutrino} species {\it weak magnetism}
their effective luminosity can be enlarged as much as 15\% compared to
the  typical one  they  achieve when  this  effect is  not taken  into
consideration during  their propagation  in dense matter.  This result
can  be interpreted  by stating  that the  number of  oscillating (and
potentially escaping) antineutrinos may be augmented by a large factor
because now  they do encounter longer  mean free paths.  Below we take
advantage of this peculiar behaviour of $\nu$ outflow in supernovae in
computing  the  overall   probability  of  transition  between  active
species and the GWs emitted in the process.



\vskip 0.5 truecm

  EDITOR PLEASE PLACE TABLE 3 AROUND HERE !!!  

\vskip 0.5 truecm

Wolfenstein (1978; 1979) and Mikheev \& Smirnov (1985) pointed out that the
neutrino oscillation pattern in  vacuum can get noticeably modified by
the passage of neutrinos through  matter because of the effect of {\it
coherent forward  scattering}. Therefore, interaction  with matter, as
pictured by  Eq.(\ref{interact}), may help in allowing  more $\nu$s to
escape  if resonant conversions  into active  (Walker \& Schramm 1987) 
and/or
sterile $\nu$s  (Mosquera Cuesta 2000) occur inside the  $\nu$-sphere 
of the
active $\nu$s.

The description of the two-neutrino  oscillations process  in  matter  
follows from  the
Schr\"odinger-like (because the dynamics is described as a function of
the space variable $x$ instead  of the standard time $t$) differential
equation (Bilenky, Giunti \& Grimus 1999; Grimus 2003)

\begin{eqnarray}
\displaystyle  i  \frac{{\rm  d}}{{\rm  d}  x}  \left(\begin{array}{c}
a_\alpha  \\ a_\beta \end{array}  \right) &  = &  \displaystyle H^{\rm
mat}_\nu   \left(\begin{array}{c}  a_\alpha  \\   a_\beta  \end{array}
\right)  \\ &  = &  \displaystyle \frac{1}{4  E}  \left\{\left[m_1^2 +
m_2^2  +  2\sqrt{2}  G_F  (  N(\nu_\alpha) +  N(\nu_\beta)  )  \right]
\left(\begin{array}{cc}  {1} & {0}  \\ {0}  & {1}  \end{array} \right)
\right.      \nonumber      \\      &      +      &      \displaystyle
\left. \left(\begin{array}{cc}  A - \Delta  m^2 \cos 2\theta  & \Delta
m^2 \cos  2\theta \\ \Delta  m^2 \cos 2\theta  & -A + \Delta  m^2 \cos
2\theta \end{array} \right) \left(\begin{array}{c} a_\alpha \\ a_\beta
\end{array} \right) \right\} \label{hamiltonian} \nonumber \; ,
\end{eqnarray}

where

\be A \equiv 2\sqrt{2} G_F E \left(N(\nu_\alpha) - N(\nu_\beta)\right) ,
\label{matter-potential}
\ee

and $  N(\nu_\alpha) \equiv \delta_{\alpha e} N_e  - \frac{1}{2} N_n$,
$\alpha, \beta  = e, \mu,  \tau, s$, $N(\nu_s)  = 0$, and  $\Delta m^2
\equiv  m_2^2 -  m_1^2$. The  eigenfunctions of  the  matter effective
Hamiltonian follow from the relation

\be H^{\rm mat}_\nu \psi_{m j} = E_j \psi_{m j}\;  ; \ee

where

\be  \psi_{m  1} =  \left(\begin{array}{c}  \cos  \theta_m  \\ -  \sin
\theta_m
\end{array} \right) , \hskip 0.5 truecm \psi_{m 2} = \left(\begin{array}{c} 
\sin \theta_m \\ \cos \theta_m \end{array} \right) .  \ee

The eigenvalues  of $E_j$ and  the matter mixing angle  $\theta_m$ are
thus given as

\begin{eqnarray}
E_{1,2} &  = & \left[m_1^2 +  m_2^2 + 2\sqrt{2} G_F  ( N(\nu_\alpha) +
N(\nu_\beta) ) \right] \nonumber \\ & \mp & \sqrt{(A - \Delta m^2 \cos
2 \theta)^2 + (\Delta m^2 \sin 2 \theta)^2 }\;,
\end{eqnarray}

and

\be \tan 2\theta_m = \frac{\tan 2\theta}{1 - \frac{A}{ \Delta m^2 \cos
2 \theta} } \label{mixing-angle}\; , \ee

where  $\theta$ is  the vacuum  mixing  angle. By  defining $U_m(x)  =
(\psi_{m 1}, ..., \psi_{m n})$ as  the mixing matrix of $n$ $\nu$, and
$\delta_j \equiv -i \int_{x_0}^{x_1} dx'$ $\psi_{m j}(x')^\dag \psi_{m
j}(x')$ as the adiabatic phases,  one can compute the {\it oscillation
amplitude} as

\begin{eqnarray}
{\mathcal{A}}^{\rm adiab}_{\nu_\alpha \rightarrow  \nu_\beta} & = & \sum_j
U_m (x_1)_{\beta j} ~~  exp \left(-i \left[\delta_j + \int_{x_0}^{x_1}
dx'    E_j(x')\right]    \right)    \nonumber    \\   &    \times    &
U^\star_m(x_0)_{\alpha j} \; . \label{oscillation-amplitude}
\end{eqnarray}

Finally,  by  averaging  over  neutrino  energies,  i.e.,  by  setting
$\left\langle  exp\left(-i \int_{x_0}^{x_1}  dx' \Delta  E(x') \right)
\right\rangle_{\rm  av} =  0$ (see  Bilenky, Giunti \& Grimus 1999),  the
outcoming transition probability among active flavor-changing species
thus reads

\be  P_{\nu_a  \rightarrow  \nu_a}(|\vec{x}_1  -  \vec{x}_0|)  =  1  -
\frac{1}{2}  \left[1  +  \cos  2\theta_m (x_0)  \cos  2\theta_m  (x_1)
\right],
\label{conversion-probability}
\ee

where  $x_0$ and  $x_1$ correspond  to the  production  (emission) and
detection sites, respectively.

\subsection{Active-to-active $\nu$ oscillations}

As stated above, in order  to produce an effect neutrinos must be able  
to escape the
core  without  thermalizing with  the  stellar  material.  For  active
neutrino species of energies $\approx 10$ MeV, this is not possible as
long  as the  matter  density  is $\geq  10^{10}$  g~cm$^{-3}$. 
Since  the
production  rate of  neutrinos  is a  steeply  increasing function  of
matter density (production rate  $\propto \rho^n$, where $\rho$ is the
matter density and $n>1$),  the overwhelming majority of the neutrinos
of all species  produced are trapped.  This way, there  seems to be no
contribution  to the  GWs  amplitude for  neutrino conversions  taking
place within the  active neutrino flavors. In the  first paper of this
series (Mosquera  Cuesta 2000), this  difficulty was overcome  only by
addressing neutrino conversions  into sterile species. Nonetheless, if
there were  indeed {\it weak  magnetism effects} (Horowitz  2002), one
can rethink  conversions within active  species. In his new  result on
{\it weak  magnetism for  antineutrinos in core  collapse supernovae},
Horowitz  (2002)     showed  that  the  antineutrinos
($\bar{\nu}_x$s) luminosity  could be noticeably  increased because of
their longer  mean free  paths, and this  means that the  total energy
flux can be  augmented in $\sim 15\%$ for  $\nu$s of temperature $\sim
10$~MeV.   One can verify  that longer  mean free  paths allows  for a
larger  oscillation probability,  and  hence the  contribution to  the
generation  of GWs  during  flavor conversions  within active  species
becomes nonnegligible  compared to  the earlier case  (Mosquera Cuesta
2000) where not weak magnetism effects were taken into account.

For  active-to-active $\nu$ flavor  conversions, for  instance: $\nu_e
\rightarrow  \nu_\mu,  \nu_\tau$;  as  implied  by  SNO  results,  the
resonance must  take place  at a distance  $x_{\rm res}$ from  the PNS
center  and  amid the  active  $\nu$-spheres,  whenever the  following
relation is satisfied (see Table \ref{matter-densities})

\be \Delta  m^2 \cos 2\theta_m = 2\sqrt{2}  ~~G_F ~~N_e(x_{\rm res})~~
k_{\nu_e} \; , \ee

notice  that  we  neglected  the magnetic  field  contribution.   Here
$k_{\nu_e}  = E_{\nu_e}/c$  is the  $\nu_e$ momentum,  and $N_e(x_{\rm
res}) =  N_{e^-} - N_{e^+} \sim 10^{39-40}$~cm$^{-3}$  is the electron
number density. Thus the right-hand part of this equation reduces to

\be 2\sqrt{2} G_F E_{\nu_e}  N_e(x_{\rm res}) = 6.9 \times 10^{4}~{\rm
eV^2}       \left[\frac{\rho}{10^{11}       \rm      g~cm^{-3}}\right]
\left(\frac{E_{\nu_e}} {\rm 10~MeV}\right) \; .  \ee

For densities of order  $\rho_{x = x_{\rm res}} \sim 10^{10}$~g~cm$^{-3}$,
i.e.,  recalling  that $\nu_e$s  are  produced  at  the PNS  outermost
regions (Walker \& Schramm 1987) where the  electron to baryon ratio is 
$Y_e
\sim  0.1$,  the resonance  condition  if  satisfied for a mass-squared
difference of  about $\Delta m^2  \sim 10^4$~eV$^2$, which  implies a
neutrino  mass  of  about  $m_\nu  \sim  10^2$~eV.   Neutrino  flavor
conversions in the resonance region  can be strong if the adiabaticity
condition is  fulfilled (Walker \& Schramm 1987), i.e.,  whenever 
(Bilenky, Giunti \& Grimus 1999; Grimus 2003)

\begin{equation}
\frac{\Delta    m^2\sin^2    2\theta_m}{2    E_{\nu}\cos    2\theta_m}
\left(\frac{1}  {\rho} \frac{{\rm  d}\rho}{{\rm  d}x}\right)^{-1}_{x =
x_{\rm res}} \gg 1 \; ,
\label{adiabaticity-condition}
\end{equation}

where $x_{\rm res}$ is the  position of the resonance layer. Recalling
that  the typical  scale  of density  variations  in the  PNS core  is
$h_{N_e}  \sim  (dN_e/dr)^{-1} \sim  $~6~km,  this adiabatic behaviour  
could be achieved as far as the density and magnetic field remain 
constant over the oscillation length

\be   
\lambda_{\rm{osc}}    \equiv   \left(\frac{1}{\rho}   
\frac{{\rm d}\rho}{{\rm d}x} \right)^{-1}_{x = x_{\rm res} }  \sim
\left(\frac{1}{2\pi} \frac{\Delta   m^2}{2  k_\nu} \sin(2\theta_m)\right)^{-1}
 \sim \frac{1~\rm cm}{\sin(2\theta_m)}\; ,
\label{oscillation-length} 
\ee

of  order $h_{N_e}$,  which  can  be satisfied  for  $\Delta m^2  \sim
10^4$~eV$^2$ as long as (Kusenko \& Segr\`e 1999)

\be \sin^2 2\theta_m > 10^{-8}\; .  \ee

Although these $\nu$ oscillations could  be adiabatic for a wide range
of mixing angles and thus a large number of $\nu$s could actually 
oscillate, a $\nu_x$ mass  such as this is incompatible with both
viable  solutions to  the Solar  Neutrino Problem  (SNP) and  the most
recent  cosmological  constraints on  the  total  mass  of all  stable
neutrino  species that  could have  left their  imprint in  the Cosmic
Microwave   Background  Radiation   (CMBR),  as   inferred   from  the
observations  performed  by the  satellite  WMAP:  $m_\nu \sim  1$~eV.
Therefore, we  dismiss this possibility  since there appears to  be no
evidence for neutrinos masses in this parameter range ($m_{\nu_x}  \sim  
10^2$~eV) inside the PNS core.

On  the other  hand, if  one takes  into account  the  KamLAND results
(Eguchi et {\it al.} 2003), which definitively demonstrated that

$i$) a large mixing angle (LMA) solution of the solar $\nu$ problem is
favoured: $\sin^2 2\theta \sim 0.8$,

$ii$)  for a  mass-squared  difference: $\Delta  m^2  \sim 5.5  \times
10^{-5}$  ~eV$^2$  (we use  the  approximate  value  $\Delta m^2  \sim
10^{-4}$~eV$^2$ for the estimates below),

one can see that resonant conversions with $\Delta m^2 = m_{\nu_2}^2 -
m_{\nu_1}^2 \sim 10^{-4}$~eV$^2$ would take place in supernova regions
where  the  density  is  about $\rho  \sim  10-30$~g~cm$^{-3}$,  which
corresponds to the outtermost layers of the exploding star. Although a
large  number  of $\nu$  species  can  in  effect participate  of  the
transitions there, i.e, the neutrino luminosity can be still a very large
quantity, these regions are  of no interest for the gravitational-wave
emission from  neutrino oscillations since the  overall energy density
at that  distance from the  star center is relatively small. This does 
not mean that no GWs are emitted from transitions there, it is to mean 
that their strain is very small so as to be detectable.  As discussed 
by Dighe \& Smirnov (2000), observations of $\nu$
oscillations in this parameter  range would provide useful information
regarding  the SNP,  the hierarchy of neutrino  masses and  the mixing
$|U_{e 3}|^2$.  Note in passing  that oscillations in this range would
imply  a  mass  for   the  $\nu_2$  species  $m_{\nu_2}  \sim 10^{-2}$
~eV ~$(\frac{\rho}{10~\rm g~cm^{-3}})$, in the case when  $m_{\nu_2} >
m_{\nu_1}$. This  $m_{\nu_2}$ is  compatible with current  limits from
WMAP (Pierce \& Murayama 2003; Hannestad 2003).

Finally, let us consider oscillations in the parameter range estimated
from CMBR by WMAP observations. In this case, resonant $\nu$ transitions
would take place in regions where the density is as high as $\rho \sim
10^{7-8}$~g~cm $^{-3}$, that is, at  the supernova mantle or PNS upper
layers.   At  these densities  the  oscillation  length  can be  still
$\lambda_{\rm  osc} \sim  1-5$~km,  and thus  the  conversions can  be
considered as adiabatic. Thus the resonance condition can be satisfied
for $\Delta m^2  \sim 1$~eV$^2$ as long as  $\sin^2 2\theta_m \leq
10^{-3}$.

Hence, by  plugging this constraint  into Eq.(\ref{mixing-angle}), and
recalling that  $i$) at least  6 $\nu$-species can participate  in the
flavor  transitions, $ii$)  most  $\nu$s are  emitted having  
$\vec{k}_\nu$ parallel to $\vec{B}$, which implies  a further reduction 
factor of 2, and  also $iii$) most $\nu$s are emitted having 
$\vec{k}_\nu$ parallel  to
$\vec{J}$ implying an additional reduction factor of  2, one can
show  from Eq.(\ref{conversion-probability}) that  the {\it{fraction}}
of  $\nu_a$ species that  can eventually  exchange flavor  during the
first   few  milliseconds   after  core   bounce  turns   out   to  be
(Mosquera Cuesta 2000)

\be  P_{\nu_a \rightarrow \nu_a}(|\vec{x}-\vec{x}_{res}|)  \sim 0.1~\%
\; .
\label{prob-active-conv}
\ee

Thus, the  total energy involved in the oscillation
process  we are considering  could be  estimated as:  $E_\nu^{\rm tot}
\simeq  P_{\nu_a  \rightarrow  \nu_a}(|\vec{x}-\vec{x}_{res}|)  \times
N_\nu  ~k_\nu  ~c ~(1  +  0.15)$, with  $N_\nu$  the  total number  of
neutrinos undergoing  flavor conversions during  the time scale $\Delta
T_{\rm osc}$. This value leads to a bit stronger  GW burst, as we show 
in Section V below.

Note  in  passing   that  KamLAND  $\bar{\nu}_e$  experiment  suggests
$P_{\bar{\nu}_e  \rightarrow  \bar{\nu}_{\mu,\tau}   }  \sim  40\%  $!
(Eguchi, et {\it al.} 2003),   while    LSND   $P_{   \bar{\nu}_\mu   
\rightarrow
\bar{\nu}_e }  \sim 0.26\% $.  This  result from LSND has  not been so
far  ruled out by  any terrestrial  experiment, and  there is  a large
expectation that it could be verified by MiniBoone at Fermilab.


\vskip 0.5 truecm

  EDITOR PLEASE PLACE TABLE 4 AROUND HERE !!!  

\vskip 0.5 truecm

\subsection{Active-to-sterile $\nu$ oscillations} 

On the other hand, if the Liquid Scintillator Neutrino Detector (LSND)
were a true indication of oscillations ($\bar{\nu}_\mu \leftrightarrow
\bar{\nu}_e$) (Pierce \& Murayama 2003; Hannestad 2003) with 
$\Delta m^2_{\rm LSND}  
\sim 10^{-1}$~eV$^2$,  then   active-to-sterile  $\nu_a  \rightarrow
\nu_s$ oscillations  could take place during a  supernova core bounce
(see Mosquera Cuesta 2002).  Such   oscillations  would  be  resonant
whenever the resonance condition is  satisfied. As seeing from Table 
\ref{matter-densities} this could happen for

\begin{equation}  
\sqrt{2}G_{F}\left[N_e(x)  - \frac{1}{2}N^0_n(x)\right] \equiv  V(x) =
\frac{\Delta m^2}{2 E_{\nu}}\cos 2\theta_m \; .  \label{resonance}
\end{equation}

Table \ref{matter-densities} also shows  that for $\nu_{\mu,\tau} 
\leftrightarrow\nu_s$ the
$N_e$ term is absent, while in the case of $\bar{\nu}$s, the potential
$  V(x)$  changes  by   an  overall  sign.   Numerically,  for  $\nu_e
\leftrightarrow \nu_s$ oscillations

\be  V(x) =  7.5 \times  10^{2} \left(\frac{\rm  eV^2}{\rm MeV}\right)
\left[\frac{\rho}{10^{10} \rm  g~cm^{-3}}\right] \left[\frac{3Y_e} {2}
- \frac{1}{2} \right]\;.  \ee

For $\nu_{\mu,\tau} \leftrightarrow\nu_s$  oscillations, the last term
in parenthesis becomes $(\frac{Y_e}{2} - \frac{1}{2}) \sim 1$.

$\nu$ conversions in the resonance  region prove to be enhanced if the
adiabaticity condition is fulfilled (Walker \& Schramm 1987). This  is 
the same as   requiring    the    oscillation      probability      in
Eq.(\ref{conversion-probability})  to   become  $P_{\nu_a  \rightarrow
\nu_s}  = \cos^2  2\theta_m \sim  1$.  Moreover,  after  the resonance
region the newly created sterile  $\nu$s have very a small probability
($<P_{\nu_s \rightarrow  \nu_a}> = \frac{1} {2}  \sin^2 2\theta_m$) of
oscillating  back  to  active   $\nu$s,  which  could  be  potentially
trapped.  It  is  easy  to  check  that  the  resonance  condition  in
Eq.(\ref{resonance}) is satisfied whenever

\be 
1~{\rm  eV}^2 \leq  \Delta  m^2_{{\nu}_a  \rightarrow {\nu}_4}
\leq 10^4~{\rm eV}^2 \; . \label{sterile-nu-mass} 
\ee

Meanwhile,            the            adiabaticity           condition,
Eq.(\ref{adiabaticity-condition}), holds if

\be  
\Delta m^2\frac{\sin^2 2\theta_m}{\cos  2\theta_m}\gg 10^{-3}{\rm
 eV^2} \left(\frac{E_{\nu}}{10\rm MeV}\right) \label{adia-num}, 
\ee

since   at  the   PNS   core  $\lambda_{\rm{osc}}   \sim  1$~km   (see
Eq.(\ref{oscillation-length})). This  is easily satisfied  for $\Delta
m^2\leq  10^4$~eV$^2$  as   long  as  $\sin^2  2\theta_m  \leq
10^{-7}$.   This  limit on  the  mixing  angle  is in  agreement  with
Superkamiokande strong constraints on $\nu_e \rightarrow \nu_s$ mixing
(Nunokawa 2001). Thence, we find  that   a
substantial  fraction ($P_{\nu_a \rightarrow \nu_s} \sim 1\%$) of $\nu$s  
may get converted to sterile $\nu$s, and  escape the core  of the star, 
if the sterile  $\nu$ mass ($m_{\nu_4}$) is such that

\be 
 1~{\rm  eV}^2  \leq  \Delta  m^2_{{\nu}_a  \rightarrow {\nu}_4}
\leq 10^4~{\rm eV}^2 \; . \label{solution}
 \ee

By looking at the lower limit for $ \Delta m^2$ in Eq.(\ref{solution}) 
one  can say  that  such a  mass difference  cannot solve the observed 
solar $\nu$ problem and/or being compatible with the atmospheric $\nu$ 
observations,  but the possibility of having three light active $\nu$s
of mass $m_{\nu_{1, 2, 3}} \sim [10^{-2}-10^{-3}]$~eV explaining these
anomalies and a ``heavy''  sterile $\nu_s$ of mass $m_{\nu_4} \leq
$~1~eV  as required  by the  LSND experiment  remains to  be  a viable
alternative. On  the other hand, if  the sterile neutrino  could be as
massive as  $m_{\nu_4} \leq$~1~keV this  mass will make it  a very
promissing  candidate   as  a  constituent  of   the  universe's  dark
matter. This  last possibility was  recently readdressed by  Fuller et
{\it al.} (2003),  who estimated sterile neutrino masses  in the range
(1-20)~keV  and small  mixing angle $\sin^2 2\theta \sim 10^{-4}$ 
with the  electron neutrino  as a
potential explanation of pulsar kicks. The same argument had also been
considered by Mosquera Cuesta (2002); and references therein.

As quoted  above, for both classes of  $\nu$ conversions the number of
$\nu$s  escaping and their  angular distribution  is sensitive  to the
instantaneous distribution of  production sites. These inhomogeneities
can   give   rise   to   quadrupole   moments   that   generates   GWs
(Mosquera Cuesta 2000; M\"{u}ller \& Janka 1997),  and dipole  moments 
that  could  drive the runaway pulsar kicks (Kusenko \& Segr\`e 1997; 
Fuller et {\it al.} 2003).  Noting that at least
6 $\nu$-species can participate in both types of oscillations and that
the interaction with  both the magnetic field and  angular momentum of
the  PNS  brings  with  an  overall  reduction  factor  of  4  in  the
oscillation  probability, one  can  show that  the {\it{fraction}}  of
$\nu$s that can  actually undergo flavor transitions in  the first few
milliseconds is (Mosquera Cuesta 2000)

\be    P_{\nu_a   \rightarrow    \nu_s}^{\nu_a    \rightarrow   \nu_a}
(|\vec{x}-\vec{x}_0|) \leq 1\% , \label{prob-conv-sterile} \ee

of  the total  $\nu$s number:  M$_\odot \times  m_p^{-1}  \sim 10^{57}
\nu$,  which  corresponds  to  a  total energy  exchanged  during  the
transition

\be \Delta  E_{\nu_a \longrightarrow \nu_s}^{\nu_a  \rightarrow \nu_a}
\sim 3\times 10^{51}~\rm erg . \label{energy-total} \ee
 
Equations  (\ref{energy-total},\ref{oscillation-time})  determine  the
total luminosity of the neutrinos participating in the resonant flavor
conversions. These  figures are called  for in the definition  used in
Eq.(\ref{nu-luminosity})   below,  as   the  basis   to   compute  the
characteristic  amplitude of  the GWs  emitted during  the transitions
discussed above.

The attentive reader must notice, however, that  the lower  limit  in
Eq.(\ref{sterile-nu-mass})  stems from  using  the constraint  derived
from  the observations  of the  CMBR
performed by the  satellite WMAP, which suggested that  the total mass
of  all  neutrino species  should  not be  larger  than  $1$~eV. In  a
hierarchy where the heaviest $\nu$  is the sterile $\nu_s$, this limit
leads to a  maximum mass: $m^{\rm max}_{\nu_s} \sim  1~{\rm eV}$.  The
use  of  this  mass-squared  difference constraint  implies  that  the
density at  the PNS  region where the  oscillations can take  place is
$\rho^{\rm   WMAP}_{m_\nu}  \sim   10^8$~g~cm$^{-3}$.   If  the   WMAP
constraint on $m_\nu$ stands up also for sterile neutrinos, this would 
preclude the mechanism for
giving  to nascent pulsars  the natal  velocities (kicks)  from active
neutrino  flavor conversions  into  sterile neutrinos,  as claimed  by
Kusenko \&  Segr\`e (1997) and Fuller  et {\it al.}   (2003), since in
such a  case the transitions would  take place in  regions far outside
the PNS core  from where no influence could be  received back once the
oscillations develop. Nonetheless, the GWs emission would still take 
place, as we stressed above.

\subsection{Experimental bounds on supernova $\nu$ oscillations}

Studies of  SN physics  have also focused  on the potential  r\^ole of
oscillations  (Walker \&  Schramm 1987)  between   active  and sterile
$\nu$s.        In    particular,     there  are     limits  on     the  
$\nu_e\leftrightarrow\nu_s$  conversion rate  inside the SN  core from 
the detected $\nu_e$ flux  from SN1987a  (Nunokawa et {\it al.}  1997; 
Nunokawa 2001).  According to  (Nunokawa et {\it al.}  1997;  Nunokawa 
2001), the time spread
and    the    number   of    detected    $\nu_e$   events    constrain
$\nu_e\leftrightarrow\nu_s$ oscillations with: $10^{6} \leq \Delta m^2
\leq  10^{8}{\rm  eV}^2$   for  $10^{-3}  \leq  \sin^2  2\theta_{\nu_e
\rightarrow  \nu_s} \leq  10^{-7}$.  More  stringent  constraints stem
from arguing that  if there were too many  ``escaping $\nu$s'', the SN
explosion  itself would  not take  place (Nunokawa et {\it al.}  1997; 
Nunokawa 2001).  Such bounds
are,  however, model  dependent.  One  should keep  in  mind that  the
mechanism through  which the  explosion takes place  is, in  fact, not
well  established (Buras,  Rampp, Janka \& Kifonidis 2003). In effect,
it has  been recently
claimed that   even after   including all  the best  physics  we  know
today about,  i.e., {\it  the state-of-the-art} on  supernova physics,
the numerical models of exploding  stars do not explode as they should
(Buras,  Rampp, Janka  \& Kifonidis 2003).  There  seems to be  a {\it 
piece} of  {\it missing} physics in those formulations.  Thence, there
seems   to  be  no   hope    of     achieving    $P_{\nu_a \rightarrow 
\nu_s}^{\nu_a \rightarrow \nu_a}$ larger  than $\sim 1$\% during $\nu$
oscillations in supernovae.

\section{GWs energetics from $\nu$ luminosity and Detectability}

If $\nu$  oscillations do  take place  in the SN  core, then  the most
likely detectable GWs signal should be produced over the time interval
for which the conditions for flavor conversions to occur are kept, i.
e., $\sim(10^{-1}-10)$~ms.  This time scale implies  GWs frequencies in
the  band: $f_{\rm{GWs}}  \sim$  [10 -  0.1]~kHz,  centered at  1~kHz,
because  of the maximum  $\nu$ production  around $1-3$~ms  after core
bounce (see  Walker \& Schramm  1987).  This frequency  range includes
the  optimal bandwidth  for detection  by  ground-based observatories.
For a 1~ms conversion time span the $\nu$ luminosity reads

\be  L_{\nu}  \equiv   \frac{\Delta  E_{\nu_a  \longrightarrow  \nu_s}
 }{\Delta  T_{\textrm{osc}}  }  \sim  \frac{3\times 10^{51}  \rm  erg}
 {1\times  10^{-3} \rm  s}  = 3  \times  10^{54} \frac{\rm  {erg}}{\rm
 s}\;. \label{nu-luminosity} \ee

Hence, the  GWs luminosity, $L_{\textrm{GWs}}$,  as a function  of the
$\nu$ luminosity can be obtained from the equation

\be 
\frac {c^3}{16 \pi G} {\left|\frac{dh}{dt}\right|}^2 = \frac {1}{4
\pi R^2} L_{\textrm{GWs}}\;, \label{GWs-luminosity}
\ee

which relates the GWs flux to  the GWs amplitude, $h_\nu$. In the case
of GWs emission  from escaping $\nu$s this amplitude  is computed from
the  expression  (Epstein 1978;  Burrows \& Hayes  1996; M\"{u}ller  \& 
Janka 1997)

\begin{eqnarray}
h_{ij}   =  \frac{2G}{c^4   R}  \int^{t   -   R/c}_{\infty}  dt^\prime
~~L_\nu(t^\prime)~~ \alpha(t^\prime)~~ e_i \otimes e_j \; ,
\label{luminosity}
\end{eqnarray}

where $e_i \otimes e_j$ represents the GWs polarization tensor.

To   attain   order  of   magnitude   estimates   one  can   transform
Eq.(\ref{luminosity}) to  get the amplitude of the  GWs burst produced
by the non-spherical outgoing front of oscillation-produced $\nu_s$ as
(Mosquera Cuesta 2000; Burrows \& Hayes 1996; M\"{u}ller \& Janka 1997)


\begin{eqnarray}
 h_\nu  & =  & \frac{2G}{c^4  R}  \left(\Delta t  \times L_\nu  \times
\alpha\right) \\ h_\nu & \simeq & |A| \left[\frac{55\rm kpc}{R}\right]
\left(\frac{\Delta  T   }{10^{-3}  \rm  s}\right)  \left[\frac{L_\nu}{
3\times     10^{54}\frac{\rm     erg}     {\rm    s}     }     \right]
\left(\frac{\alpha}{0.1}\right) \; , \label{burrows96a} \nonumber 
\end{eqnarray}


where $|A| = 4 \times 10^{-23}~ {\rm Hz}^{-1/2}$ is the amplitude.
\footnote{One must  realize at this point  that the high  value of the
anisotropy  parameter  here  used  is consistently  supported  by  the
discussion regarding  the neutrino  coupling to rotation  and magnetic
field   presented  above.}  Equivalently,   that  amplitude   can  be
reparametrized as

\be  
h_\nu \simeq  |A|  \left[\frac{55\rm kpc}{R}\right]  \left(\frac{
\Delta   E_{\rm    GWs}}{10^{-7}\rm   M_\odot   c^{2}}   \right)^{1/2}
\left[\frac{10^{3}\rm                    Hz}{                   f_{\rm
GWs}}\right]^{1/2}~\label{amplitude}\;.  
\ee

A  GWs  signal  this strong  will  likely  be  detected by  the  first
generation of GWs interferometers as  LIGO, VIRGO, etc. Its imprint in
the GWs waveform may resemble  a spike of high amplitude and timewidth
of  $\sim$~ms followed  by a  Christodoulou's memory  (Mosquera Cuesta
2002).  From Eq.(\ref{burrows96a}) the GWs luminosity turns out to be

\be 
L_{\textrm{GWs}}  \sim  10^{50-48} ~~  \frac{\rm {erg}}{\rm s}
~~   \left[\frac{L_\nu}{3\times  10^{54}   \frac{\rm   {erg}}{\rm  s}}
\right]^2 \left(\frac{\alpha}{0.1-0.01}\right)^2\;, 
\ee


 while the GWs energy radiated in the process yields

\be    E_{\textrm{GWs}}   \equiv   L_{\textrm{GWs}}    \times   \Delta
T_{\textrm{osc}} \sim  10^{47-45}~{\rm erg}\;.  \ee

 This  is about $10^{5-3}  \times L^{\rm  Diff.}_\nu$ the  luminosity from
$\nu$  diffusion  inside the  PNS,  as  estimated earlier (Mosquera Cuesta
2000; 2002).


\vskip 0.5 truecm

  EDITOR PLEASE PLACE TABLE 5 AROUND HERE !!!  

\vskip 0.5 truecm

\vskip 0.7 truecm

\section{Conclusion} 

One can see that if  $\nu$ flavor conversions indeed take place during
SN  core bounce  inasmuch as  they  take place  in our  Sun and  Earth
(Smirnov 2002), then GWs should be released during the transition.
The GWs  signal from  the process is  expected to irradiate  much more
energy than current mechanisms figured out to drive the NS dynamics at
birth do. A luminosity  this large (Eq.(\ref{GWs-luminosity})) would turn
these bursts the strongest GWs signal  to be detected from any SN that
may come to occur, futurely, on  distances up to the VIRGO cluster, $R
\sim [10-20]$ Mpc.  It is stressed  that this signal will still be the
stronger one  from a  given SN, even  in the  worst case in  which the
probability of  $\nu$ conversion is three orders  of magnitude smaller
then the estimated  in the present paper. In proviso,  we argue that a
GWs signal  that strong could  have been detected during  SN1987a from
the  Tarantula  Nebula  in  the  Milky Way's  satellite  galaxy  Large
Magellanic Cloud, despite  of the low sensitivity of  the detectors at
the epoch. In such a case,  the GWs burst must have been correlated in
time with the earliest arriving neutrino burst constituted of some active
species given  off during the  very early oscillation  transient where
some  $\nu_e$s  went   into  $\nu_{\mu}$s,  $\nu_\tau$s  or  $\nu_s$s.
Thenceforth, it could be of  worth to reanalyze the data collected for
from that  event taking careful follow  up of their  arrival times, if
appropriate timing was available at that moment.

\section{Acknowledgements}

{The authors truly thank Profs. I. Dami\~ao Soares and  R.  Zukanovich
Funchal for their patience in reviewing this manuscript 
and their valious criticisms and suggestions. HJMC acknowledges the
support from the Funda\c c\~ao de Amparo \`a Pesquisa do Estado do Rio
de Janeiro (FAPERJ), Brazil, through the grant E-26/151.684/2002. KF
thanks CAPES (Brazil) for a graduate fellowship,  Prof. Sandra D. 
Prado (IF-UFRGS) for continued support and advice, and ICRA-BR (Rio 
de Janeiro) for hospitality and support during part of this work.}

\newpage

EDITOR: THIS IS FIGURE 1    !!!!

\begin{figure}[ht]
\centerline{\epsfxsize=3.5truein\epsffile{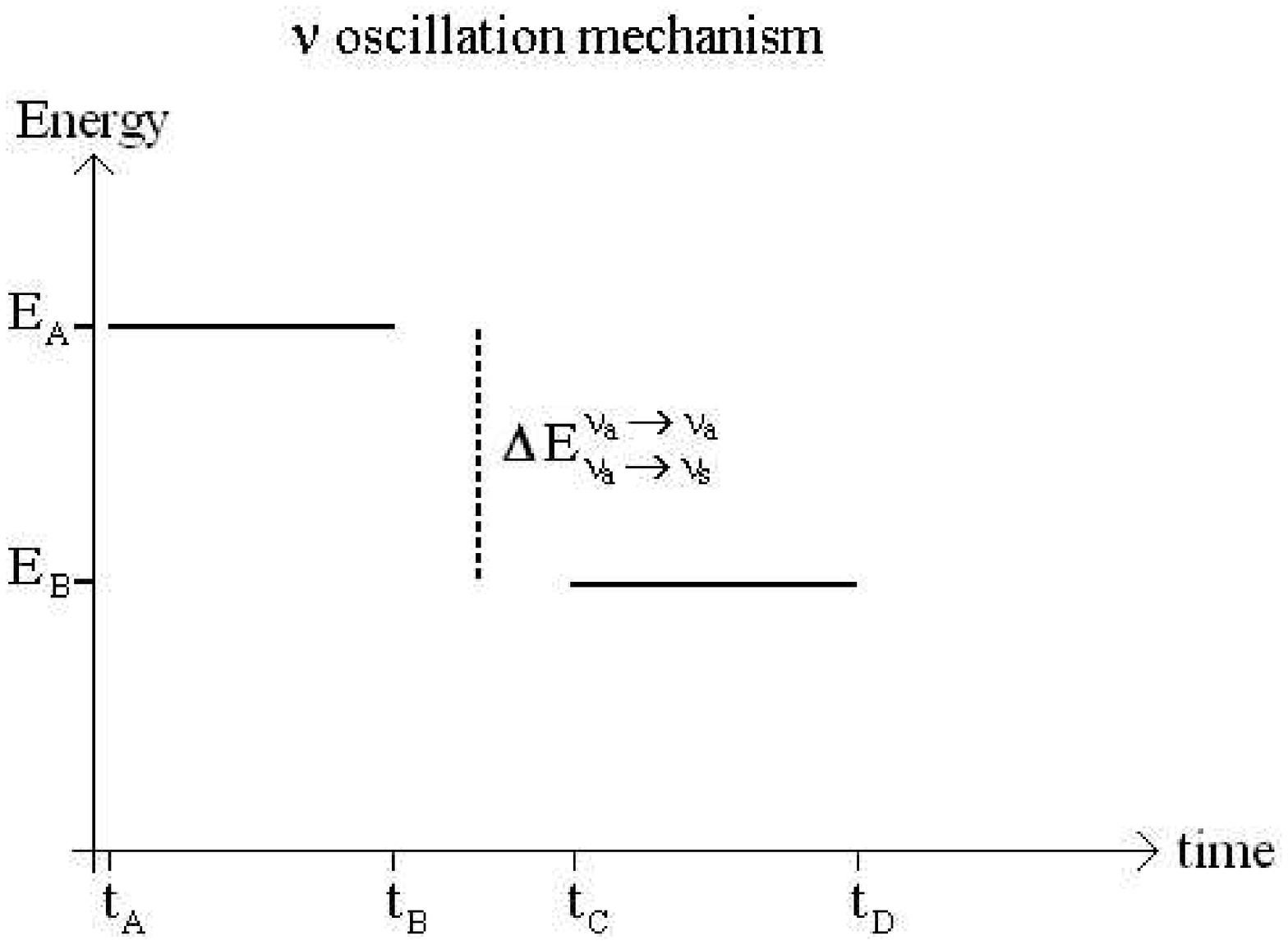}}
\caption{The mechanism for $\nu$  conversions. Firstly, the state of a
system, $\nu$s in the PNS, is defined by the energy $E_A$ for the time
interval $t_B  - t_A$.  After the flavor  transitions that  system is
represented by a reduced energy $E_B$ over time $t_D - t_C$. The total
energy   transferred  to   the   new  species   is  $\Delta   E_{\nu_a
\leftrightarrow \nu_a}^{\nu_a \leftrightarrow \nu_s} \equiv E_A - E_B$,
and thus the $\nu$ luminosity over the transient is given by $L_\nu = 
\frac{ E_A - E_B}{ t_C - t_B}$.}
\end{figure}

\newpage

EDITOR: THIS IS FIGURE 2    !!!!

\begin{figure}[ht]
\centerline{\epsfxsize=3.5truein\epsffile{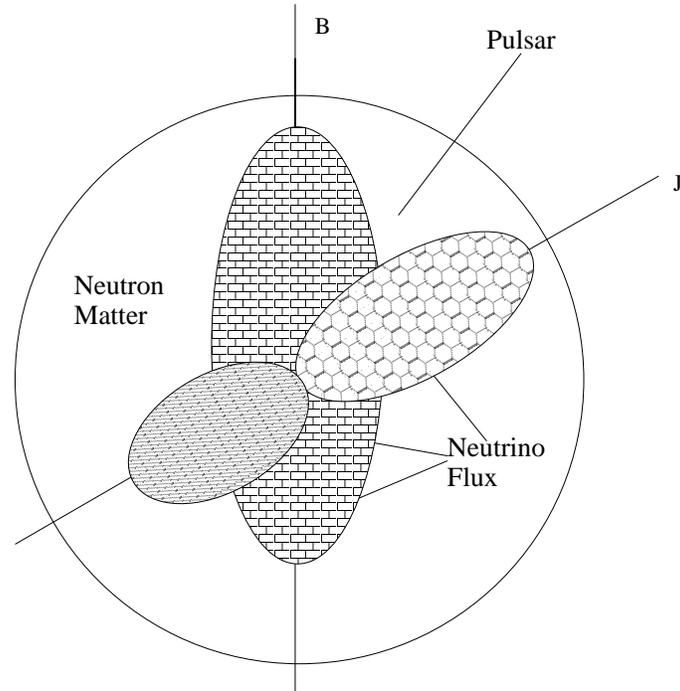}}
\caption{A  schematic representation  of  the $\nu$-flux 
distribution  (hatched regions)  in  a nascent  pulsar.  The at  least
quadrupolar  distribution  is  evident  and stems  from  the  neutrino
spin-magnetic and spin-rotation couplings.}\label{4-dupole}
\end{figure}

\newpage

EDITOR: THIS IS TABLE 1    !!!!



\begin{table}[bth]
\caption{Time  scales {(\rm in  ms)} called for in   this    paper.}  
\begin{tabular}{cccc}
\multicolumn{1}{c} {   } & 
\multicolumn{1}{c} {   } & 
\multicolumn{1}{c} {   } &
\multicolumn{1}{c} {   } \\  
\vspace{3pt} 
{Supernova    $\nu$} & {Neutrino } & {Supernova} & { Supernova } \\
{Thermalizat.} & { Oscillations } & {Deleptonizat.  } & { Core-Bounce }  \\
$\Delta  T_{\nu}^{\rm thermal} $  & $  \Delta T_{\rm  osc}^{\nu} $ & $\Delta T_{\rm delept}^{\rm PNS} $  & $ \Delta T_{\rm bounce}^{\rm PNS}$ \\ 
$  \sim 500 $ & $  \sim 3-10 $ & $  \sim 10 $ & $ \sim  20 $ \\
\end{tabular}  \label{time-scales}
\end{table}

\newpage

EDITOR: THIS IS TABLE 2    !!!!

\begin{table}
\caption{Parameters of the PNS used for in this paper.}
\begin{tabular}{ccccc}  
\multicolumn{1}{c}{ } & \multicolumn{1}{c}{ } & \multicolumn{1}{c}{{} } &
\multicolumn{1}{c}{ } & \multicolumn{1}{c}{ } \\
{Radius  } & {  Mass   }  &  {Density }  & { $\nu$-sphere } & {Ang. Mom. } \\
{$  R_{\rm  PNS} $ } &  {$ M_{\rm  PNS}$}  & {$\rho_{\rm  PNS}$}  &  { $ R_{\rm  PNS}^{\nu}$ } &  {  $a$ }   \\
$ \sim 80\rm ~km $  & $\sim  1.3~\rm M_\odot $ & $ 3\times 10^{11}\rm g~cm^{-3}$ & {$ \sim 35-60 \rm ~km $} & { $\sim 0.9-1 $ } \\ 
{ } & { } & { $2\times 10^{14}\rm g~cm^{-3}$ } & { } & { } \\
\end{tabular}  \label{parameters-PNS}
\end{table}

\newpage

EDITOR: THIS IS TABLE 3   !!!!

\begin{table}{ccccc}
\caption{Matter densities relevant for two-$\nu$ oscillations. {\bf Notice} that  $\nu_\mu \rightarrow \nu_\tau$  oscillations take place as  in vacuum.  The coupling Fermi  constant is: $G_F^2  = 5.29 \times 10^{-44}$~cm$^2$ MeV$^{-2}$, or  equivalently   $G_F  =  10^{-49}$~erg~cm$^{-3}$. } 
\begin{tabular}{ccccc}  
\multicolumn{1}{c}{ } & \multicolumn{1}{c}{ } & \multicolumn{1}{c}{{} } &
\multicolumn{1}{c}{ } & \multicolumn{1}{c}{ } \\ 
 { }  & { $\nu_e \rightarrow \nu_{\mu, \tau} $  } & { $\nu_e \rightarrow \nu_s$
}  & {$\nu_\mu \rightarrow  \nu_\tau$} &  { $\nu_{\mu, \tau} \rightarrow \nu_s $ } \\
$\frac{A}{2\sqrt{2} G_F E}$ & $N_e$  & $N_e -  \frac{1}{2} N_n$  & $0$  & $-  \frac{1}{2} N_n  $ \\
\end{tabular} \label{matter-densities}
\end{table}

\newpage 

EDITOR: THIS IS TABLE 4   !!!!

\begin{table}
\caption{Parameter range for GWs emission from $\nu$s oscillations in SNe. 
The symbol DM stands for $\nu_s$ as a  Dark Matter candidate.}
\begin{tabular}{ccccccc}  
\multicolumn{1}{c}{ } & \multicolumn{1}{c}{ } & \multicolumn{1}{c}{ } &
\multicolumn{1}{c}{ } & \multicolumn{1}{c}{ } & \multicolumn{1}{c}{ } &
\multicolumn{1}{c}{ } \\
\vspace{3pt}  
{ $\nu$-\rm  sphere  } &  {$\nu$  Luminosity }  & { Oscil. Length } & {$\nu$ Velocity } & { $\Delta m^2 ({\nu_a\leftrightarrow \nu_a})$ } &
{  $\Delta m^2({\nu_a\leftrightarrow \nu_s})  $ } & 
{  $\Delta m^2({\nu_a\leftrightarrow \nu_s})  $ }  \\
{  $R_{\rm PNS}^{\nu_e}$  } &  { $L({\nu-Diff.})$  } &
{ $\lambda_{\rm osc}^{\nu}$} & {$\bar{V}_{\nu-Diff.}$ } & 
{ (SNO,KamLAND) } & { (WMAP) } & { ($\nu_s$~DM) }  \\  
$\sim 35$~km  & $\sim 10^{53}\rm  ~erg~s^{-1}$ &  $\sim 1\rm ~km$  & $  
\sim 10^{9}\rm ~cm~s^{-1}  $  &  $\sim  10^{-4}\rm  ~eV^2$ &  $  \sim  
1\rm  ~eV^2$ & $  \sim  10^4 \rm ~eV^2$ \\ 
\end{tabular} 
\end{table}

\newpage 

EDITOR: THIS IS TABLE 5   !!!!

\begin{table}  
\caption{GWs estimates from $\nu$ oscillations and diffusion.}
\begin{tabular}{cccc} 
\multicolumn{1}{c}{ } & \multicolumn{1}{c}{ } & \multicolumn{1}{c}{{} } &
\multicolumn{1}{c}{ } \\  
{ GWs Freq.   } &  { GWs  Energy } & { $L_\nu$ Asymm. } &  {GWs Energy } \\
{$f_{\rm  GW_s} $} & 
{ $E_{\rm GW}(_{\nu_a \leftrightarrow \nu_a}^{\nu_a\leftrightarrow \nu_s})$ } 
& { $\alpha$ } &  {  $E_{\rm GW}({\nu-Diff.})$ } \\
$[10-0.1]\rm  ~kHz $ &  $[10^{-6}-10^{-8}]\rm ~M_\odot~c^2$  & $[0.1-0.01]$ & 
$ [10^{-10}-10^{-13}] \rm ~M_\odot~c^2 $ \\
\end{tabular} 
\end{table}

\end{document}